\documentstyle[12pt]{article}
\newcommand{\beq}{\begin{equation}}
\newcommand{\eeq}{\end{equation}}
\newcommand{\beqa}{\begin{eqnarray}}
\newcommand{\eeqa}{\end{eqnarray}}
\newcommand{\beqar}{\begin{eqnarray*}}
\newcommand{\eeqar}{\end{eqnarray*}}

\begin{document}
\begin{titlepage}
\vspace{.5in}
\thispagestyle{empty}
\begin{center}
{\bf\Large  
Quantum Measurement Back-Reaction and  Induced  Topological Phases }

\vspace{.4in}

{\large Y. Aharonov$^{(a,b)}$, 
T. Kaufherr$^{(a)}$\footnote{trka@ccsg.tau.ac.il}, 
S. Popescu$^{(c)}$\footnote{\it sp230@newton.cam.ac.uk}
, and B. Reznik$^{(d)}$\footnote{\it reznik@t6-serv.lanl.gov\\
LAUR-97-1588}}\\
\medskip
{(a) \small \it School of Physics and Astronomy, 
Tel-Aviv University, Tel Aviv 69978, Israel} \\
\medskip
{{(b)} \small \it Department of Physics,
University of South Carolina, Columbia, SC 29208}\\
\medskip
{(c) \small \it Isaac Newton Institute, Cambridge University, 
 20 Clarkson Rd., Cambridge, CB3 OEH, U.K., 
and BRIMS, Hewlett-Packard Labs., Bristol BS12 6QZ, U.K.}\\
\medskip
(d) {\it \small Theoretical Division, T-6, MS B288, 
Los Alamos National Laboratory, Los Alamos, NM, 87545}
\end{center}
\vspace{.5in}
\begin{center}
\begin{minipage}{5in}
\begin{center}
{\large\bf Abstract}
\end{center}
{\small
It is shown that a topological vector-potential (Berry phase) is induced 
by the act of measuring angular momentum in a direction
defined by a reference particle. 
This vector potential appears as a consequence
of the back-reaction due to the quantum measurement.
}
\end{minipage}
\end{center}
\end{titlepage}

As is well-known, many of the most common observables (position, velocity, 
angular momentum, etc..), 
both in classical mechanics as well as in quantum 
mechanics, are {\it relative} observables -- 
they always are defined relative to 
a system of reference \cite{aharonov-susskind1,aharonov-kaufherr}. 
Indeed, we never measure the absolute position of a particle, 
but the distance in between the particle and some other object; 
similarly, we never measure the angular 
momentum of a particle along an absolute axis, but along a direction defined by 
some other physical objects. Anything can constitute a ``reference system", from 
macroscopic bodies to microscopic particles, but they are always there, even if, 
for simplicity, we don't always refer to them explicitly. 
Obviously, measuring a system relative to a frame of reference implies an 
interaction in between the system and the reference system (via the measuring 
apparatus), and thus affects both. 
It is here that the quantum mechanical case differs 
considerably from  the classical case. 
The uncertainty principle implies that unlike classical mechanics, 
the  quantum mechanical back-reaction can never be negligible.

In this letter we show that, 
the ``strong'' nature of the quantum mechanical
back-reaction on the reference system, can  in particular
cases give rise to an effective topological vector-potential.
This induced topological effect can  
be interpreted as a Berry phase, thus leading to a fundamental 
relation between quantum measurements and the Berry effect. 

To begin with, let us consider a measurement 
of a half-integer spin $\vec s$ in the direction defined by a quantum particle 
of mass M.  In other words, we consider the measurement of the observable $\vec 
s_{\hat n} = \hat n \cdot \vec s$ where $\vec s$ is the spin and $\hat n=\vec 
r/|\vec r|$ is the direction of the reference particle as seen in the laboratory 
frame of reference.
Choosing  the reference particle to be free (except for the coupling with the 
spin during the measurement), and the measuring interaction to be 
von-Neumann-like, the total Hamiltonian is:
\beq
H = {\vec P^2\over 2M} + H_s + g(t)q \hat n \cdot \vec s .
\label{measure}
\eeq
Here $H_s$ stands for the Hamiltonian of the spin system. 
The measurement is described by the last term:  $q$ is
a canonical variable of the measuring device and its conjugate, $P_q$, plays the 
role of the ``pointer";   $g(t)$ is a time
dependent coupling constant, 
which we shall take to satisfy  $\int g(t) dt = 1$. 
For the special case of a constant coupling, $g(t)=1/T$ for $0<t<T$ and 
zero otherwise, the shift of the position of the pointer yields the average 
relative spin:
$P_q(T)-P_q(0)= {1\over T}\int_0^T \hat n \cdot \vec s dt$.   
In the limiting case  of $T\to 0$, we obtain 
$g(t)\to \delta(t)$, which  corresponds to the ordinary von-Neumann measurement.  

Notice that in the limit of a continuous measurement, for which $g(t)=constant$
in a finite time interval, the von-Neumann interaction term in Eq.
(\ref{measure}), has the same form as the well known monopole-like example of a
Berry phase \cite{berry-phase}.  Thus a Berry phase is expected upon rotation of
the reference system.  But while in Berry's case the interaction is put in ``by
hand" just to study its consequences, in our case the interaction naturally
arises whenever the spin is measured.

As we shall see, the appearance of the Berry phase and the associated vector
potential, can be easily obtained by transforming to a quantum reference-frame.
There the spin observable becomes directly measurable and the back-reaction felt
by the reference particle is precisely given by the requisite Berry vector
potential.

Let us consider first the  2-dimensional case. 
The reference axis is given in terms of the unit vector 
$\hat n = \hat x\cos\phi + \hat y \sin\phi$, 
and $\hat n \cdot \vec s = s_x\cos\phi + s_y\sin\phi$. (Here $\hat x$ and $\hat 
y$  and $\phi$ denote the standard coordinate unit vectors and respectively the 
polar angle in the laboratory frame of reference.)
The last term in the Hamiltonian (1) above, can be simplified
by transforming to a new set of variables.
The unitary transformation: 
\beq
U_{(2d)} = e^{-i\phi(s_z - {1\over2})}  
\label{u2d}
\eeq
yields the relations:
\beq
p'_\phi  = U^\dagger p_\phi U = p_\phi - (s_z - 1/2) \ \ ;  \ \ p_r' = p_r 
\ \ ; \ \ \vec r\ ' = \vec r,
\eeq
\beq
 s_x' = s_x\cos\phi + s_y \sin\phi \ \ ; \ \ 
s_y' = s_y\cos\phi - s_x\sin\phi \ \ ; \ \ 
s_z' = s_z.
\eeq
The effect of this unitary transformation is to define new spin variables and a 
new canonical momentum for the reference particle, while the coordinates of the 
reference particle (defined  with respect to the laboratory) remain unchanged.
The extra $1/2$ factor in (\ref{u2d}) is required in order 
to preserve the single valueness of the wave function,
of the combined spin and reference particle system,  
with respect to the angular coordinate $\phi$. (For an integer
spin we  drop the $1/2$).

Expressing the Hamiltonian in terms of the new 
variables we obtain:
\beq
H = {1\over 2M}\biggl(  \vec p\ ' +  {s_z' -1/2\over r} 
\hat\phi  \biggr)^2 + H'_s + g(t)q s_x' .
\eeq
In these variables, 
the measuring device interacts directly with $s_x'$. The relative spin
$s_x'$ is a measurable observable, which commutes with the 
total angular momentum, $p'_\phi-1/2$,  since $[s_x', p_\phi']=0$. 
We notice however that in the new variables 
the reference particle sees the  effective vector potential
\beq
\vec A_{(2d)} = {s_z' -  {1/2}\over r}\hat\phi .
\eeq
The latter describes the back-reaction on
the reference frame, which here takes the form of a 
fictitious magnetic  
fluxon at the origin $\vec r=0$,  
with a magnetic flux $\Phi = s_z' - 1/2$ in the $\hat z$-direction.
 In the absence of the measurement, ($g(t)=0$), the  $s'_z$ component of the 
spin is
a constant of motion.
Thus, the $2s+1$  components of the
 wave function in the $s_z'$ representation evolve independently. 
The vector potential corresponding to the
$s'_z=m_s$ component is  $\vec A_{(2d)}=(m_s-1/2)\hat\phi/r$, 
i.e. it corresponds to an  integer number of quantized fluxons.
Since for all the components the 
vector potential is equivalent to a pure gauge transformation, 
it causes no observable effect on the reference particle. 
On the other hand, during the measurement, the interaction 
with the measuring device causes a rotation of $s_z'$, which in turn leads to 
observable effects. The rotation of $s_z'$ and the exact character of the 
associated effects depends on the relative strength of the different terms in 
(5). 
In the present work we are interested in the limit of ``ideal", (i.e. very 
accurate) measurements. In this limit the interaction hamiltonian dominates all 
other terms. Indeed, in order for the measurement to be accurate, the initial 
position of the pointer, $P_q(0)$ must be precisely known, i.e. $\Delta 
P_q(0)\rightarrow 0$. In turn, this implies that the uncertainty in $q$ is very 
big, $\Delta q\geq 1/\Delta P_q(0)\rightarrow \infty$, that is, the typical 
values of $q$ in the interaction hamiltonian are infinite. As a consequence, in 
this limiting case the spin components $s_y'$ and $s_z'$, which are
orthogonal to $\hat n$, rotate with infinite frequency, and can be averaged to 
zero. (In the original variables (1) the spin is a ``fast" degree of freedom 
which follows adiabatically the slow motion of the reference particle.) 

More exactly, the typical frequency of rotation of the spin components $s_y'$
and $s_z'$, associated with the interaction hamiltonian is $\omega_s \approx
g/\Delta P_q= 1/(T\Delta P_q)$.  This is to be compared with the frequency
associated with $H'_S$, the ``free" hamiltonian of the spin, and with the
frequency associated to the kinetic term.  The later one is the more important
as it scales at least as $1/T$; indeed, to see the Berry phase one needs to
perform an interference experiment with the reference particle during the time
of the measurement, i.e.  the duration of the interference experiment 
$T_{exp}\leq T$.  When the ratio of the angular frequencies is
$\omega_s/\omega_r \approx T_{exp}/(T\Delta P_q) >> 1$ (which is always reached 
when the precision of the measurement is increased while keeping all other 
parameters constant) we are in the adiabatic regime.

In the adiabatic regime corresponding to an ideal measurement the effective 
vector potential seen by the 
reference system can therefore be
obtained by taking the expectation value of 
$\vec A_{(2d)}$ with respect to the spin wave function:
\beq
\langle \vec A_{(2d)}  \rangle = 
\langle {s_z-1/2\over r }\hat\phi \rangle 
\approx {1/2\over r} \hat \phi . 
\eeq
This corresponds to a semi-quantized fluxon at the origin 
$r=0$,  pointing to the  $\hat z$ direction.
The total phase accumulated in a cyclic motion of the reference
particle around the semi-fluxon yields the topological 
(path independent) phase:
\beq
\gamma_n = \oint A_{(2d})dl = n\pi ,
\label{semi-flux}
\eeq
where $n$ is the winding number.

Note that in the above case the exact values of $g$ and $\Delta P_q$ are
essentially irrelevant - all that is needed is for them to be such that the
adiabatic regime holds.  On the other hand, 
outside the adiabatic regime, 
the interaction term does not completely dominate the
other terms, the exact values of $g$ and $\Delta P_q$ become essential, and the
consequences of the measurement are much more complicated; this case is outside 
of our present interest. 

Consider now the  case,  of a free reference particle in 
3-dimensions, the appropriate transformation which maps:
$s_x'= U_{(3d)}^\dagger s_x U_{(3d)} = \hat n\cdot \vec s$ is
\beq
U_{(3d)} = e^{-i(\theta-\pi/2)s_y} e^{-i\phi(s_z-1/2)},
\eeq
where $\theta$ and $\phi$ are spherical angles.
\cite{note}

The corresponding 3-dimensional vector-potential is in this case 
\beq
A_{(3d)x} = -s_y {\cos\theta\cos\phi\over r} +
                  \Bigl( s_z \sin\theta + s_x\cos\theta -{1/2}
           \Bigr){\sin\phi\over r\sin\theta},
\eeq
\beq
A_{(3d)y} = -s_y {\cos\theta\sin\phi\over r} -
                  \Bigl( s_z \sin\theta + s_x\cos\theta -{1/2}
           \Bigr){\cos\phi\over r\sin\theta},
\eeq
\beq
A_{(3d)z} = s_y {\sin\theta\over r}.
\eeq
For the case of an integer spin or angular momentum 
the $1/2$ above is omitted.
It can be verified that  $\vec A_{(3d)}$ corresponds to a pure 
gauge non-Abelian vector potential. 
The field strength vanishes locally, 
$F_{\mu\nu}= \partial_\mu A_\nu  - \partial_\nu A_\mu 
-[A_\mu, A_\nu]=0$. 
Thus the force on the reference particle vanishes.
Furthermore since the loop integral, 
$\oint \vec A_{(3d}) \cdot d\vec r$,
gives rise to a trivial flux $2n\pi$, the manifold is simply
connected.  
(This can be seen by noticing that the 
 magnetic field, $\nabla \times \vec A_{(3d)}$,
due to the terms proportional to $s_y$ vanishes. The  
other terms correspond to a 
fluxon pointing in the $\hat z$-direction with total flux
$\Phi=  s_z \sin\theta + s_x\cos\theta -{1/2}$  which is quantized
for spin components along the direction $\pi/2 -\theta$.) 
Thus, as in the 2-d case, in absence of coupling with  the 
measuring device, $\vec A_{(3d)}$  is a pure gauge vector potential.

In the adiabatic limit discussed above, during the measurement 
we have $\langle  s_z\rangle  \approx 
\langle s_y \rangle \approx 0$. The effective vector 
potential seen by the reference particle
\beq
\langle \vec A_{(3d)} \rangle=  \Bigl(s_x \cos\theta - 1/2\Bigr)
{\hat \phi\over r\sin \theta },
\eeq
is identical to the (asymptotical, $r\to \infty$)     
non-Abelian 't Hooft - Polyakov monopole 
\cite{monopole} in the unitary gauge.
The effective magnetic field, $\nabla \times \langle 
\vec A_{(3d)}\rangle$:
\beq
\langle \vec B\rangle = s_x {\vec r\over r^3}
\eeq
corresponds to that of a magnetic monopole 
at the origin, $r=0$, with a magnetic charge $m= s_x$. 

The topological vector potential obtained above,  clearly have an
observable manifestation.
Upon rotation of the reference particle around the $\hat z$ axis, 
the particle accumulates
an Aharonov-Bohm phase:
\beq
\gamma_n = \oint \vec A_{(3d)}\cdot d\vec r = 
- n \pi ( 1-  \cos \theta),
\eeq
which equals  half of the solid angle subtended by the path.
The latter can be observable by means of a standard 
interference experiment. We thus conclude that during  a
continuous measurement the back-reaction on the reference particle
takes the form of a topological vector potential, 
of a semi-fluxon in 2-dimensions and that of a monopole 
in 3-dimensions.

Our discussion above can also be restated  in terms 
of Berry phases. 
Viewing the reference particle as
a slowly changing environment, 
and the spin system as a fast system which is driven by a 
time dependent `environment',
we can use the Born-Oppenheimer procedure 
to solve for the spin's eigenstates. Let us assume for simplicity
that $g_0$ is sufficiently large so $H_s$ can be neglected,
and that $g(t)$ is roughly constant.
Considering for simplicity the 2-dimensional case, 
the appropriate eigenstate equation therefore reads
\beq
g q \hat n(\phi) \cdot \vec s |\psi(\phi)\rangle = 
E |\psi(\phi)\rangle , 
\eeq
where $\phi$ is here viewed as the  external parameter. 
For simplicity let us consider the case of $s=1/2$. 
We obtain:
\beq
|\psi_\pm(\phi)\rangle = {1\over\sqrt2}\Bigl( 
e^{-i\phi/2}|\uparrow_z\rangle \pm e^{+i\phi/2}|\downarrow_z\rangle
\Bigr)
\otimes|q\rangle .
\eeq
The eigenfunctions $|\psi_\pm\rangle$ are double-valued in 
the angle $\phi$.  
Thus a cyclic motion in space, which changes $\phi$
by $2\pi$, induces a sign change. The latter
is due to the `spinorial nature' of fermionic particles, 
which as is well known, flips sign under a $2\pi$ rotation 
\cite{dirac}. 
To obtain the appropriate Berry phase we 
need to construct single valued solutions
of Eq. (17): 
\beq
|\Psi(\phi)\rangle = e^{-i\phi(s_z+ 1/2)}|\uparrow_x\rangle
= \Bigl( 
e^{-i\phi}|\uparrow_z\rangle \pm  |\downarrow_z\rangle
\Bigr) \otimes|q\rangle .
\eeq
It then follows that the Berry phase \cite{berry-phase}:
\beq
\gamma_{Berry} = i\oint \langle \Psi(\phi)|{\partial_\phi\over r}
|\Psi(\phi)\rangle d\phi  = \gamma_n,
\eeq
is identical to the phase, (\ref{semi-flux}), which is 
induced by the effective semi-fluxon. Similarly, the Berry phase
in the 3-dimensional  case corresponds to half of the  solid angle 
subtended by the path of the reference particle.

In conclusion, we have shown that the quantum mechanical back-reaction 
during a measurement induces in certain cases a topological
vector potential. The Berry phase can be viewed in this 
framework, as a necessary consequence of the ``strong'' nature
of the quantum back-reaction. 

 \vspace {2 cm} 

{\bf Acknowledgment}
Y. A. achnowledges the support of  the Basic
Research Foundation,  grant 614/95, 
administered by the Israel Academy of Sciences and Humanities.


\begin{thebibliography}{99}
\bibitem{note}
In the operator $e^{-i\phi(s_z-1/2)}$ describing the rotation by the spherical 
angle 
$\phi$, a factor of $-1/2$
 is added to the spin operator $s_z$ in order to insure the
 single-valuedness of the transformed wave-function.
 On the other hand no such factor needs to be added 
to the spin operator $s_y$ in the corresponding operator 
$e^{-i(\theta-\pi/2)s_y}$ which describes the rotation by the 
$\theta-\pi/2$ spherical angle. The reason is that while the angle $\phi$ takes 
values
 in between $[0, 2\pi]$, the angle $\theta$ takes values only in between 
$[0,\pi]$.
\end{thebibliography}
\end{document}